\documentclass[aps,pra,superscriptaddress,showkeys,showpacs,twocolumn,floatfix]{revtex4}

\usepackage{amssymb,amsbsy,amsmath}

\usepackage{graphicx}
\usepackage{dcolumn}
\usepackage{bm}
\usepackage{subfigure}
\usepackage{color}

\newcommand{\op}[1]{%
    \fontdimen12\textfont3=2pt\fontdimen12\scriptfont3=1.4pt%
    \!\null\mathop{\vphantom{#1}\smash{#1}}\limits_{\sim}\null\!}
\newcommand{\xref}[1]{\protect\ref{#1}}
\newcommand{\figref}[1]{Fig.~\protect\ref{#1}}
\newcommand{\fmref}[1]{(\protect\ref{#1})}
\def\bra#1{\langle \, {#1} \, | \,}
\def\ket#1{\, | \, {#1} \, \rangle}
\newcommand{\braket}[2]{\langle \, {#1} \, | \, {#2} \, \rangle}
\newcommand{\pp}[2]{\frac{\partial \, {#1}}{\partial \, {#2}}\;}
\newcommand{\Tr}{\mbox{Tr}}
\begin{document}
\title{Advanced Finite-Temperature Lanczos Method for
  anisotropic spin systems}

\author{Oliver Hanebaum}
\author{J\"urgen Schnack}
\email{jschnack@uni-bielefeld.de}
\affiliation{Fakult\"at f\"ur Physik, Universit\"at Bielefeld, Postfach 100131, D-33501 Bielefeld, Germany}

\date{\today}

\begin{abstract}
It is virtually impossible to evaluate the magnetic properties of
large anisotropic magnetic molecules numerically exactly due to
the huge Hilbert space 
dimensions as well as due to the absence of symmetries. Here we
propose to advance the Finite-Temperature Lanczos Method (FTLM) to the
case of single-ion anisotropy. The main obstacle, namely the
loss of the spin rotational symmetry about the field axis, can
be overcome by choosing symmetry related random 
vectors for the approximate evaluation of the partition
function. We demonstrate that now thermodynamic functions
for anisotropic magnetic molecules of unprecedented size can be
evaluated. 
\end{abstract}

\pacs{75.10.Jm,75.50.Xx,75.40.Mg} \keywords{Heisenberg
model, Magnetic molecules, Numerical Approximation, Magnetization}

\maketitle

\section{Introduction}
\label{sec-1}

The magnetism of anisotropic spin systems in particular magnetic
molecules is very rich and leads to interesting phenomena such as
bistability and quantum tunneling, both related to the
anisotropy barrier \cite{GSV:2006}. Nevertheless many Single
Molecule Magnets (SMM) such as Mn$_{12}$ acetate
\cite{Lis:ACB80,STS:JACS93,SGC:Nat93,TLB:Nature96,GNS:PRB98,CAG:JMMM01,GaS:ACIE03} 
constitute a massive challenge for theoretical investigations
since the underlying Hilbert space of the molecular many-spin
system is orders of magnitude too big for an exact and complete
matrix diagonalization. In cases such as that of Mn$_{12}$ the
lowest zero-field split multiplet is largely separated from the
rest of the energy spectrum so that it can approximately be
treated as a single giant spin for low-enough temperatures.

For molecules where such a separation is impossible one would
like the investigate the full spectrum in order to understand
their magnetic properties. The largest species where such a procedure was
possible are the Mn$_6^{\text{III}}$ of Euan Brechin's group
\cite{MVW:JACS07,CGS:PRL08} as well as the Mn$_6^{\text{III}}$M
molecules of Thorsten Glaser's group
\cite{GHK:IC09,Gla:CC11,HHK:DT12,HHK:IC12,HKH::EJIC13,HSB:IC14}. 
The numerically most demanding member of the latter family,
Mn$_6^{\text{III}}$Cr$^{\text{III}}$, possesses a Hilbert space
dimension of 62.500, which thanks to inversion symmetry can be
reduced to half the size. Nevertheless, a single calculation for
one external magnetic field value including a powder average over
25 directions needs about a week on an 8 core workstation, not
to mention the necessary 46~GB of RAM. Exact calculations for larger
magnetic molecules are thus virtually impossible. It would therefore
be very appealing to have a reliable approximation at one's
disposal. 

In the realm of spin systems that are described by the
Heisenberg model the Finite-Temperature Lanczos Method
(FTLM) \cite{PhysRevB.49.5065,JaP:AP00,MHL:CPL01,LPE:PRB03,ADE:PRB03}, which is a
so-called trace estimator, was successfully applied to
temperature and field dependent magnetic observables for 
lattice systems \cite{SSP:EPJB04,ZST:PRB06,STS:PRB07,SSZ:NJP12},
to optical conductivities \cite{JaP:PRB94} and a
variety of molecules \cite{ScW:EPJB10,ScH:EPJB13} up to very
large Hilbert spaces with dimensions of the order of $10^{10}$
\cite{ZZL:CC13}. Although the underlying Lanczos method
\cite{Lan:JRNBS50} is not restricted to isotropic spin models
and has also been used to determine low-lying eigenstates of
Mn$_{12}$ acetate \cite{RJS:PRB02,CSO:PRB04}, the accuracy of
FTLM largely increases if symmetries of the Hamiltonian can be
exploited. In anisotropic spin systems the spin-rotational,
i.e. SU(2) symmetry or even the simpler $\op{S}^z$-symmetry are
lost. Therefore, a straight forward extension of FTLM appears
doubtful. 

In this article we demonstrate that by restoring time-reversal
invariance in the set of initial random vectors used for FTLM
the accuracy of magnetic observables can be drastically improved
compared to a naive ansatz. Interestingly, this mainly concerns
high-temperature quantities such as $\chi T$ vs. $T$ or
$\mu_{\text{eff}}$ vs. $T$, which without restoring symmetry
tend to systematically deviate from the correct result, i.e. the
paramagnetic limit. We show with a few 
examples that FTLM yields results that are virtually
indistinguishable from the exact ones and that one can now treat
systems of unprecedented size, which is exemplarily demonstrated
for a 
fictitious Mn$_{12}^{\text{III}}$ molecule.

The article is organized as follows. In Section~\xref{sec-2}
basics of the Finite-Temperature Lanczos method are
repeated. Section~\xref{sec-3} discusses the problem of reduced
symmetries.  Section~\xref{sec-4} discusses the performance of
the method for 
large molecules. The article closes with summary and outlook.

\section{Recapitulation of the Finite-Temperature Lanczos Method}
\label{sec-2}

The exact partition function $Z$ depending on
temperature $T$ and magnetic field $B$ is given by a trace
\begin{eqnarray}
\label{E-1-1}
Z(T,B)
&=&
\sum_{\nu}\;
\bra{\nu} e^{-\beta \op{H}} \ket{\nu}
\ ,
\end{eqnarray}
where $\{\ket{\nu}\}$ denotes an orthonormal basis of the
respective Hilbert space. Following the ideas of
Refs.~\cite{PhysRevB.49.5065,JaP:AP00} the unknown matrix
elements are approximated as
\begin{eqnarray}
\label{E-1-2}
\bra{\nu} e^{-\beta \op{H}} \ket{\nu}
&\approx&
\sum_{n=1}^{N_L}\;
\braket{\nu}{n(\nu)} e^{-\beta \epsilon_n^{(\nu)}} \braket{n(\nu)}{\nu}
\ .
\end{eqnarray}
For the evaluation of the right hand side of Eq.~\fmref{E-1-2}
$\ket{\nu}$ is taken as the initial vector of a Lanczos
iteration of $N_L$ steps, which generates a respective Krylov
space. As common for the Lanczos method 
the Hamiltonian is diagonalized in this Krylow space, which
yields the $N_L$ Lanczos eigenvectors $\ket{n(\nu)}$ as well as
the associated Lanczos energies $\epsilon_n^{(\nu)}$, where
$n=1,\dots, N_L$. The notation $n(\nu)$ reminds one that the 
$\ket{n(\nu)}$ belong to the Krylov space
derived from the original state $\ket{\nu}$.

The parameter $N_L$ needs to be large enough to reach the
extremal energy eigenvalues but should not be too large in order
not to run into problems of numerical accuracy. $N_L\approx 100$
is a typical and good value.

In addition, the complete and thus very large sum over all
states $\ket{\nu}$ is 
replaced by a summation over a set of $R$ random
vectors. The partition function is thus approximated by 
\begin{eqnarray}
\label{E-1-3}
Z(T,B)
&\approx&
\frac{\text{dim}({\mathcal H})}{R}
\sum_{\nu=1}^R\;
\sum_{n=1}^{N_L}\;
e^{-\beta \epsilon_n^{(\nu)}} |\braket{n(\nu)}{\nu}|^2
\ .
\end{eqnarray}
Symmetries can be taken into account by applying the procedure
for every orthogonal subspace ${\mathcal H}(\Gamma)$, i.e.
\begin{eqnarray}
\label{E-1-4}
Z(T,B)
&\approx&
\sum_{\Gamma}\;
\frac{\text{dim}({\mathcal H}(\Gamma))}{R_{\Gamma}}
\sum_{\nu=1}^{R_{\Gamma}}\;
\sum_{n=1}^{N_L}\;
\nonumber \\
&&\times
e^{-\beta \epsilon_n^{(\nu,\Gamma)}} |\braket{n(\nu,
  \Gamma)}{\nu, \Gamma}|^2 
\ .
\end{eqnarray}
$\Gamma$ denotes the irreducible representations of the
symmetry group. Observables are then evaluated as
\begin{eqnarray}
\label{E-1-5}
O(T,B)
&\approx&
\frac{1}{Z(T,B)}
\sum_{\Gamma}\;
\frac{\text{dim}({\mathcal H}(\Gamma))}{R_{\Gamma}}
\sum_{\nu=1}^{R_{\Gamma}}\;
\sum_{n=1}^{N_L}\;
e^{-\beta \epsilon_n^{(\nu,\Gamma)}}
\nonumber \\
&&\times
\bra{n(\nu, \Gamma)}\op{O}\ket{\nu, \Gamma}
\braket{\nu, \Gamma}{n(\nu, \Gamma)}
\ .
\end{eqnarray}
The very positive experience is that even for large problems the
number of random starting vectors as well as the number of
Lanczos steps can be chosen rather small, e.g. $R\approx 100,
N_L\approx 100$ \cite{ScW:EPJB10,ScH:EPJB13}. Since Lanczos
iterations consist of matrix vector multiplications they can be
parallelized by \verb§openMP§ directives. In our programs
this is further accelerated by an analytical state coding and an
evaluation of matrix elements of the Hamiltonian ``on
the fly" \cite{SHS:JCP07}.

\section{The problem of anisotropic spin systems}
\label{sec-3}

For the anisotropic spin systems considered in this publication
the complete Hamiltonian of the spin system is given by the
Heisenberg term, the single-ion anisotropy, and the Zeeman term, 
i. e.  
\begin{eqnarray}
\label{E-3-1}
\op{H}
&=&
-
2\;
\sum_{i<j}\;
{J}_{ij}
\op{\vec{s}}_i \cdot \op{\vec{s}}_j
+
\sum_{i}\;
\op{\vec{s}}_i \cdot 
{\mathbf D}_i
\cdot \op{\vec{s}}_i
\\
&&+
\mu_B\, B\,
\sum_{i}\;
g_i
\op{s}^z_i
\nonumber
\ .
\end{eqnarray}
${J}_{ij}$ is the exchange parameter between spins at sites $i$
and $j$. A negative ${J}_{ij}$ corresponds to an
antiferromagnetic interaction, a positive one to a ferromagnetic
interaction. For the sake of simplicity
it is assumed that the $g_i$ are numbers.
${\mathbf D}_i$ denotes the single-ion anisotropy tensor, which in
its eigensystem $\vec{e}_{i}^{\,1}$, $\vec{e}_{i}^{\,2}$,
$\vec{e}_{i}^{\,3}$, can be decomposed as
\begin{eqnarray}
\label{E-4-8}
{\mathbf D}_i
&=&
D_i \vec{e}_{i}^{\,3} \otimes \vec{e}_{i}^{\,3}
+
E_i
 \left\{
\vec{e}_{i}^{\,1} \otimes \vec{e}_{i}^{\,1}
- 
\vec{e}_{i}^{\,2} \otimes \vec{e}_{i}^{\,2}
\right\}\ .
\end{eqnarray}
The  magnetization can be derived from the thermodynamic
potential $G(T,\vec{B})$
\begin{eqnarray}
\label{E-3-4-A}
\vec{{\mathcal M}}(T,\vec{B})
&=&
-
\pp{}{\vec{B}}\, G(T,\vec{B})
\\
\label{E-3-4-B}
G(T,\vec{B})
&=&
-k_B T \
\text{ln}\left[Z(T,\vec{B})\right]
\ .
\end{eqnarray}
In the following we consider only the spatial component of
$\vec{{\mathcal M}}$ that is parallel to the field direction. 
To avoid the very costly evaluation of eigenvectors we
approximate the derivative in \fmref{E-3-4-A}, equivalently the
evaluation of the magnetization according to \fmref{E-1-5}, by a
difference quotient. 

The major problem of a straight forward application of FTLM is
the general loss  of symmetries, and in view of the
magnetization the loss of the $\op{S}^z$-symmetry.
To understand this aspect better we would like to repeat the
benefits of an $\op{S}^z$-symmetry. This symmetry means that the
complete Hilbert space can be decomposed into mutually
orthogonal subspaces ${\mathcal H}(M)$ for each total magnetic
quantum number $M$. For each energy eigenvalue in ${\mathcal
  H}(M)$ there exists a degenerate eigenvalue in ${\mathcal
  H}(-M)$. Therefore, with a Lanczos procedure one would only
generate the approximate levels for non-negative $M$ and take those
for negative $M$ as copies, which automatically preserves the
$\op{S}^z$-symmetry in the pseudo spectrum. In
terms of the magnetization this guarantees the very general
symmetry
\begin{eqnarray}
\label{E-3-5}
\vec{{\mathcal M}}(T,-\vec{B})
&=&
-
\vec{{\mathcal M}}(T,\vec{B})
\ .
\end{eqnarray}
It is also related to the properties of magnetic observables at
high temperatures since these rely on trace formulas such as
\begin{eqnarray}
\label{E-3-6}
\Tr\left(\op{S}^z\right)
&=&
0
\ ,
\end{eqnarray}
as can be seen in high-temperature expansions
\cite{SSL:PRB01,TWB:DT10,SLR:PRB11,LSR:PRB14}. If a relation
such as \fmref{E-3-6} 
is violated in an approximation the high-temperature limit of
the magnetization (or susceptibility) does not correspond to the
correct paramagnetic limit.

In an approximation which rests on random states, as FTLM
does, a symmetry that is broken by every random vector, is only
restored in the limit of very large sets of random realizations
(central limit theorem). A scheme such as outlined above, where
one duplicates every Lanczos energy eigenvalue for the subspace
with negative magnetic quantum number, restores the
$\op{S}^z$-symmetry even for small numbers of 
random vectors. For anisotropic spin systems such a scheme
is not applicable, because the simple  $\op{S}^z$-symmetry does no
longer apply. But the 
very general symmetry \fmref{E-3-5}, that goes back to time
reversal invariance of Hamiltonian \fmref{E-3-1} when the
magnetic field is inverted simultaneously, still applies. It
means, that every Lanczos energy eigenvector that is evaluated
for a certain field $\vec{B}$ has a time-symmetric counterpart
that is the respective eigenvector for $-\vec{B}$. Taking
$\vec{B}$ as the quantization direction,
i.e. $\vec{B}=B\vec{e}_z$, this yields for a Lanczos energy
eigenvector 
\begin{eqnarray}
\label{E-3-7}
\ket{n(\nu)}
&=&
\sum_{\vec{m}}
\, c_{\vec{m}} \ket{\vec{m}}
\end{eqnarray}
the following vector as symmetry-related counterpart
\begin{eqnarray}
\label{E-3-8}
\ket{\tilde{n}(\nu)}
&=&
\sum_{\vec{m}}
\, c_{\vec{m}}^* \ket{-\vec{m}}
\ .
\end{eqnarray}
Here $\ket{\vec{m}}$ is a state of the product basis (where each
single-spin state is an eigenstate of the single $\op{s}_i^z$
operator), while 
$\ket{-\vec{m}}$ denotes the basis state, where all single
magnetic quantum numbers $m_i$ are inverted compared to $\ket{\vec{m}}$. The
coefficients $c_{\vec{m}}^*$ are complex conjugated with respect
to $c_{\vec{m}}$. It is very important
to note, that the two states are not degenerate. $\ket{n(\nu)}$
is an eigenstate of \fmref{E-3-1} for $B$ with magnetization
${\mathcal M}_n=\bra{n(\nu)}\mu_B\sum_{i}g_i\op{s}^z_i\ket{n(\nu)}$, 
whereas $\ket{\tilde{n}(\nu)}$ is an eigenstate for $-B$, but
then indeed with the same energy and opposite magnetization,
i.e. $-{\mathcal M}_n$. 

It turns out that the application of such a procedure would be
very costly, since all eigenstates \fmref{E-3-7} would be needed
in order
to construct the symmetry-related states \fmref{E-3-8}, for
which the energy expectation value (at negative $B$) would have
to be evaluated. We therefore propose the following
simplification. For every random starting vector $\ket{\nu}$ of
our Lanczos procedure we also take its symmetry-related
counterpart as a random starting vector. This does not exactly
guarantee \fmref{E-3-6}, but comes close to a rather high
precision.

\section{Application to large spin systems}
\label{sec-4}

\begin{figure}[ht!]
\centering
\includegraphics*[clip,width=30mm]{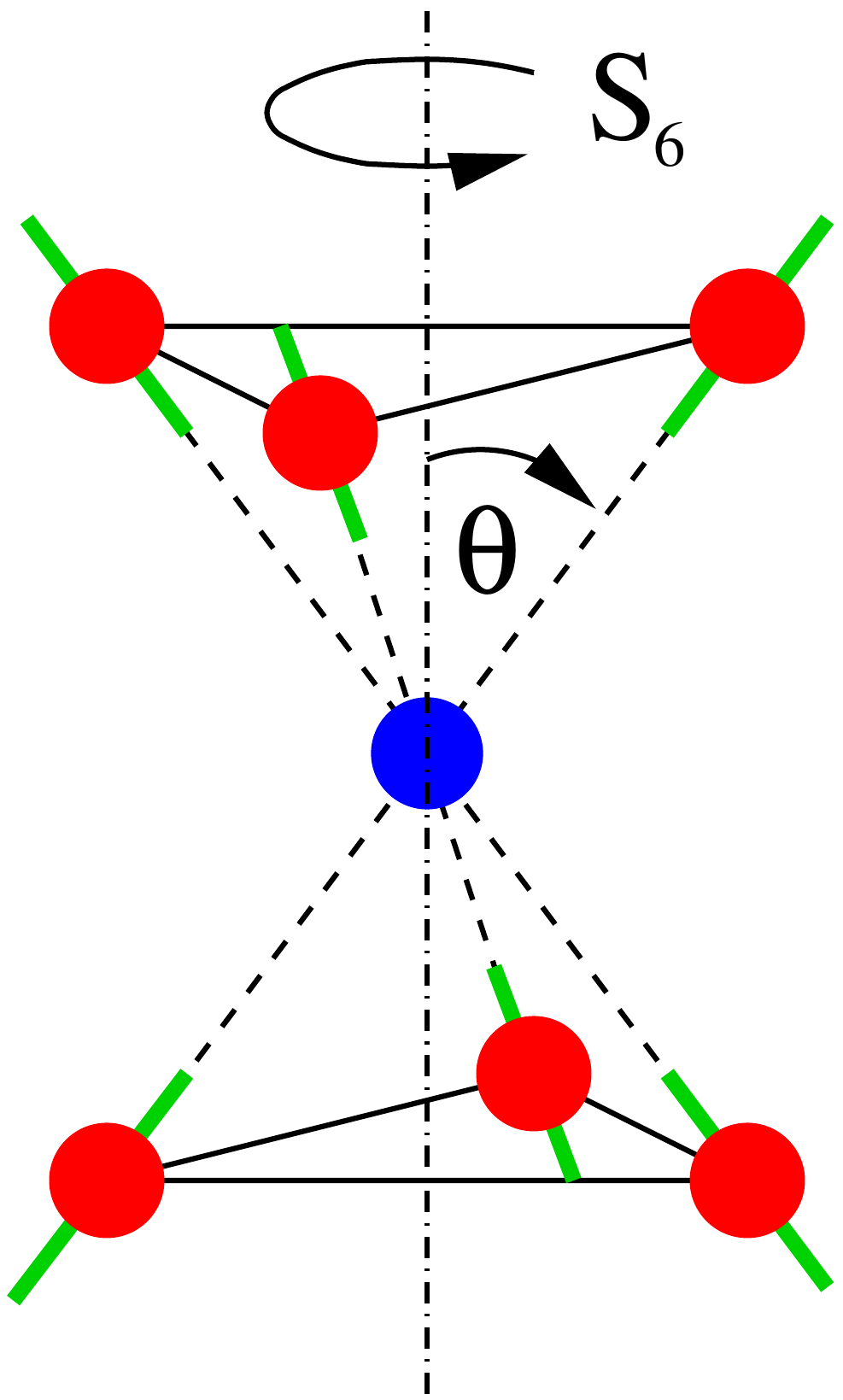}
\caption{Structure of the S$_6$-symmetric molecules \textbf{M1}
  and \textbf{M2} investigated in this section, compare also
  Ref.~\cite{Gla:CC11}. \textbf{M1}: 6 spins in two triangles
  without central spin; \textbf{M2}: like \textbf{M1} but with
  seventh, central spin. The thick short bars denote easy axes.}  
\label{tlmm-f-1}
\end{figure}

In the following we demonstrate the potential of the method. In
the first part we choose model systems that can
also be treated exactly since this allows to estimate the
numerical accuracy qualitatively without having to care about
experimental 
uncertainties or inappropriate parameters of the model.
Model system \textbf{M1} is inspired by Mn$_6^{\text{III}}$
molecules \cite{GHT:DT10} with 6 spins $s=2$ arranged in two 
uncoupled equilateral triangles. We choose a fictitious nearest
neighbor exchange interaction $J=\pm 0.314$~cm$^{-1}$ and
single-ion anisotropy tensors with $D_i=-5.0$~cm$^{-1}$, $E_i=0$
and an angle of $\Theta=40^\circ$ of the local easy axis to the
S$_6$-symmetry axis of the molecule. The polar angles differ by 
$\Delta\phi=120^\circ$ between neighbors according to the
S$_6$-symmetry, compare \figref{tlmm-f-1} without central ion.  

\begin{figure}[ht!]
\centering
\includegraphics*[clip,width=70mm]{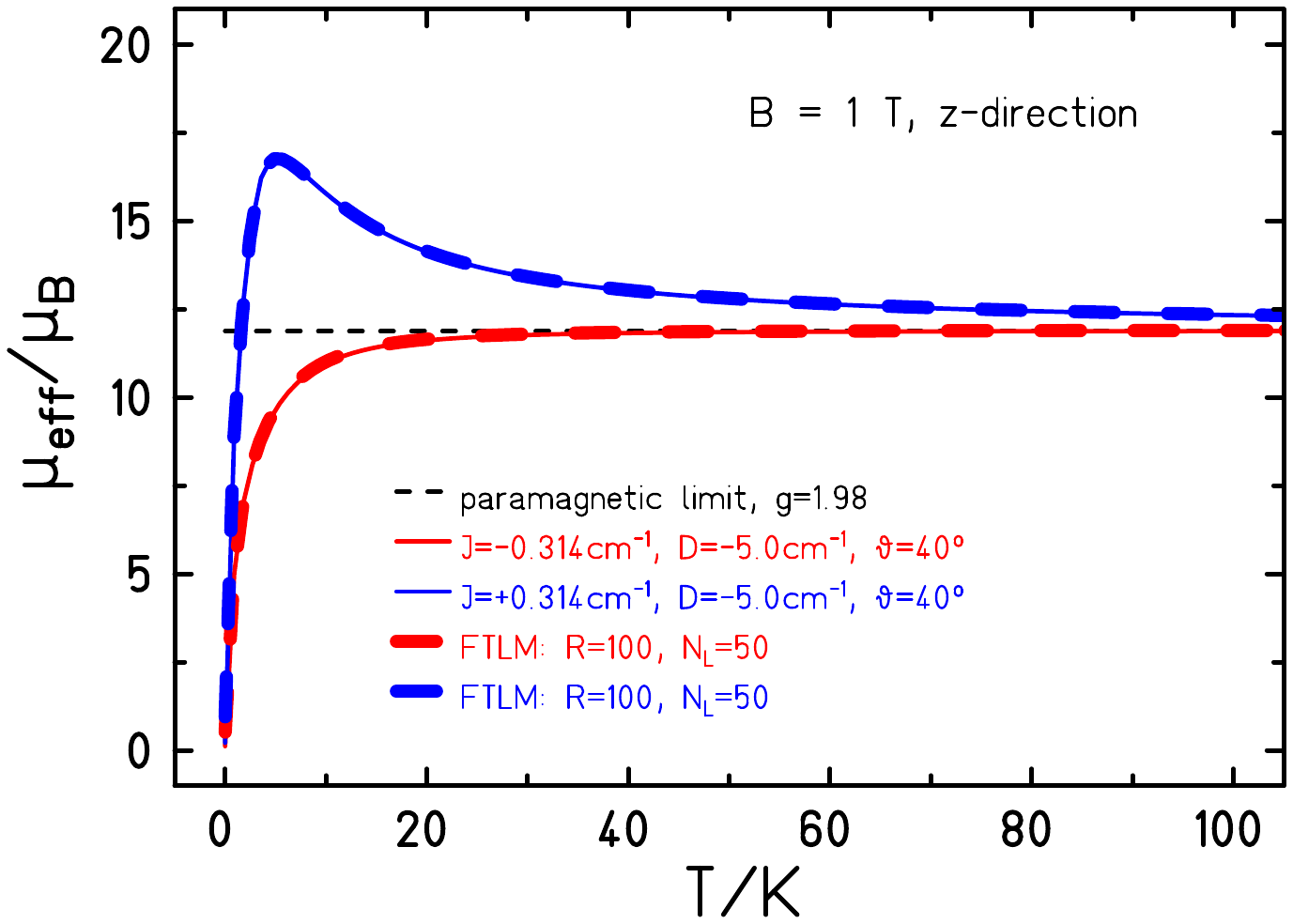}

\includegraphics*[clip,width=70mm]{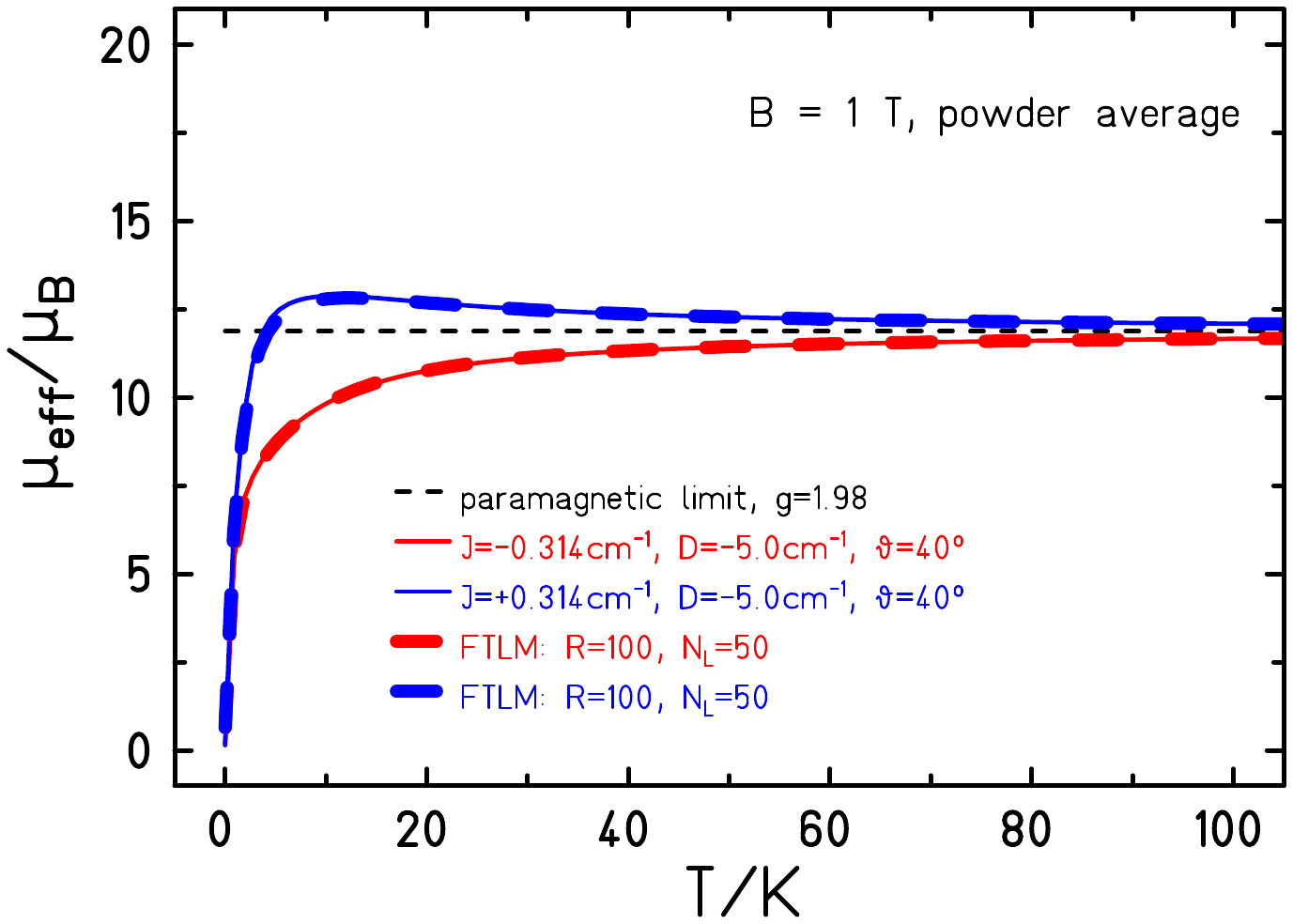}
\caption{Effective magnetic moment of model system \textbf{M1}:
  moment along $z$-direction (top), powder averaged
  moment (bottom). The solid curves show the result of full
  matrix diagonalization, the dashed ones the result of FTLM.}   
\label{tlmm-f-2}
\end{figure}

Figure \xref{tlmm-f-2} shows the effective magnetic moment of
model system \textbf{M1} along $z$-direction (top) and as a
powder average (bottom). The solid curves show the result of full
matrix diagonalization, the dashed ones the result of FTLM. The
powder average was performed using a Lebedev-Laikov grid of 50
orientations \cite{LeL:DAN99}. For the FTLM we used
100 random vectors together with their respective
symmetry-related counterparts and just 50 Lanczos steps. As was
observed in other FTLM simulations the results are very good,
only sometimes a tiny deviation is observed for temperatures of
the order of typical parameters of the Hamiltonian. Note that
the low-temperature properties are bound to be very accurate
since low-lying states are approached exponentially fast with the
number of Lanczos steps. For this reason the low-temperature
magnetization is not shown in this section.

\begin{figure}[ht!]
\centering
\includegraphics*[clip,width=70mm]{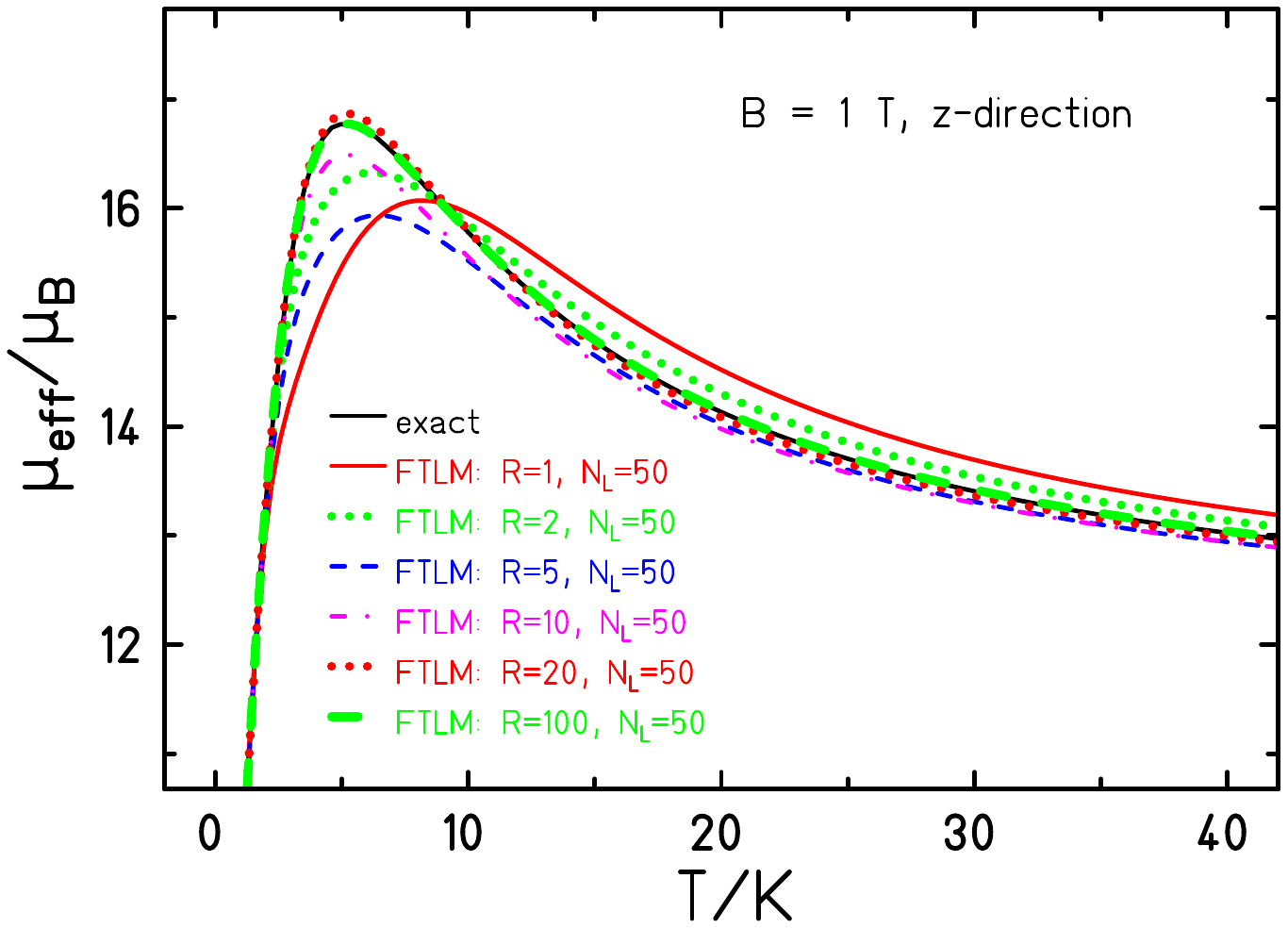}

\includegraphics*[clip,width=70mm]{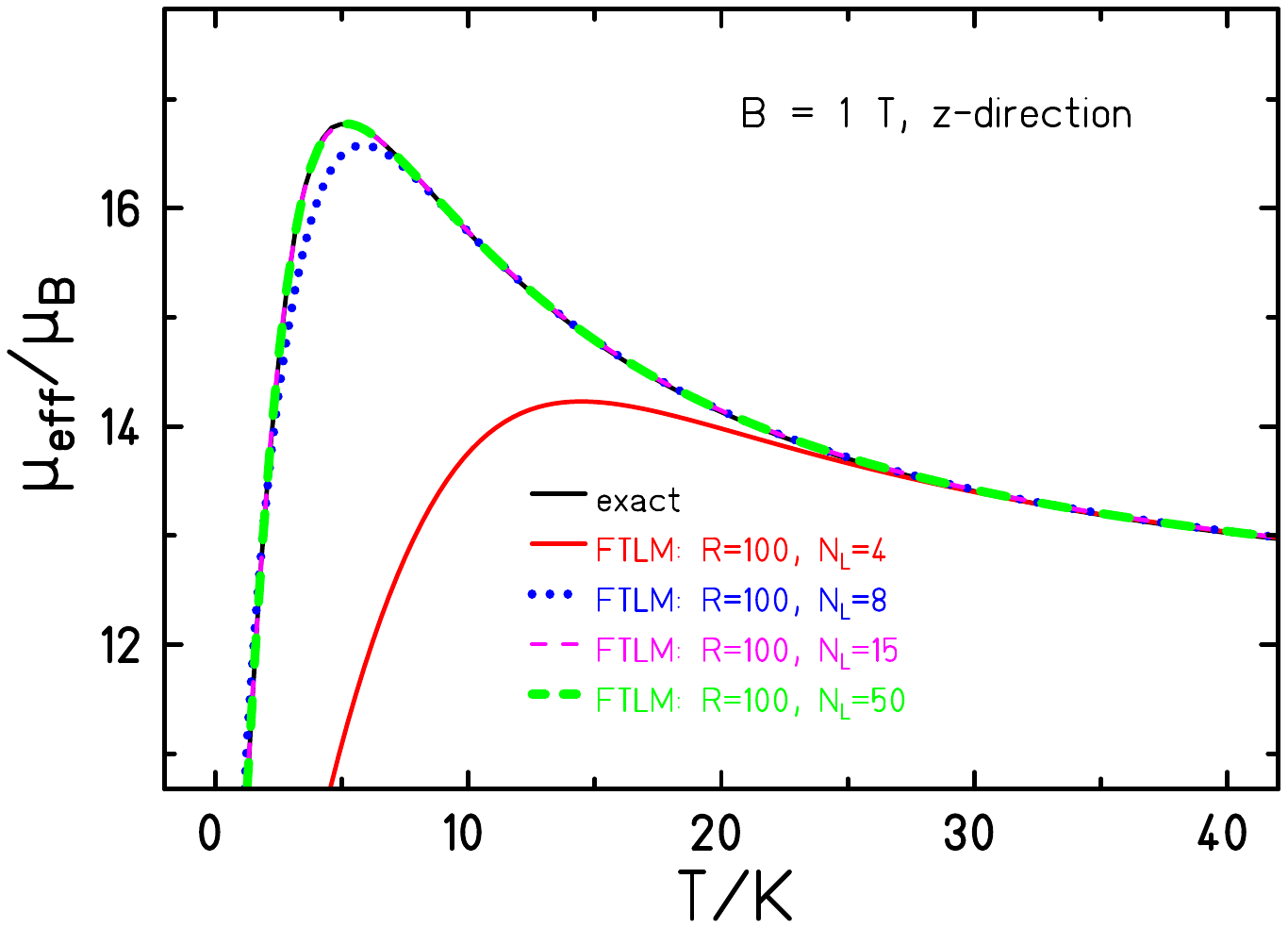}
\caption{Effective magnetic moment of model system \textbf{M1}:
  The solid curves show the result of full matrix 
  diagonalization, the broken curves show results of FTLM with
  reduced numbers of random vectors (top) and Lanczos steps
  (bottom).} 
\label{tlmm-f-3}
\end{figure}

In order to gain some insight into the influence of the two
parameters $R$ and $N_L$ of the approximation, we
performed simulations with a few different values. Figure
\xref{tlmm-f-3} (top) shows how the approximation approaches the
exact result as a function of $R$ for fixed $N_L=50$. One
notices that even for a small number of random vectors low- and
high-temperature part are already rather accurate, and that for
an overall convergence $R\gtrsim 20$ (plus symmetry-related
states) is already sufficient. The approximation for $R=100$ is
indistinguishable from the exact result. The bottom part of
\figref {tlmm-f-3} displays FTLM approximations for a few $N_L$
with fixed number of random vectors $R=100$. It is amazing how
quickly the approximation approaches the exact result: for 
$N_L\gtrsim 15$ no deviation is visible any more.

\begin{figure}[ht!]
\centering
\includegraphics*[clip,width=70mm]{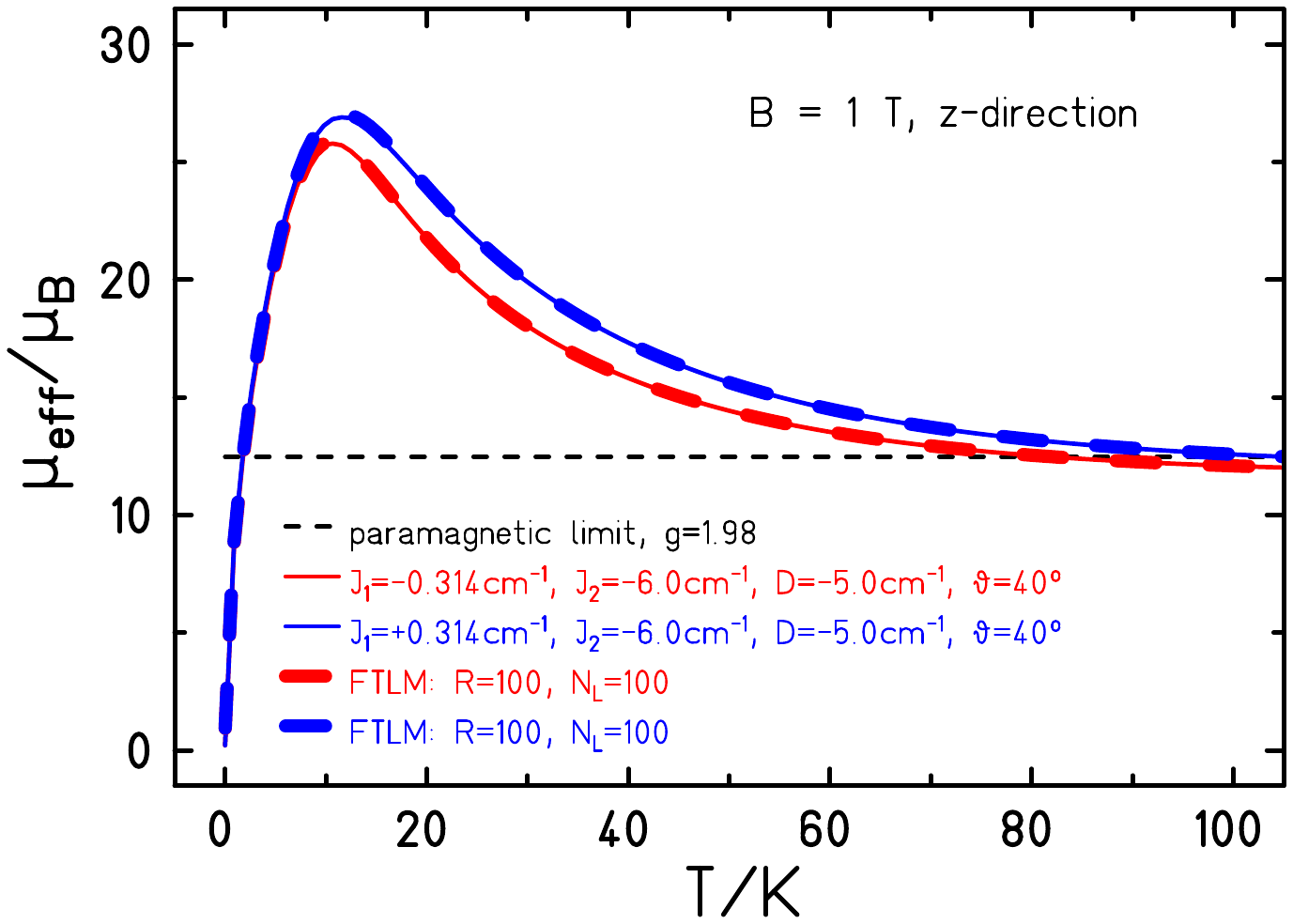}

\includegraphics*[clip,width=70mm]{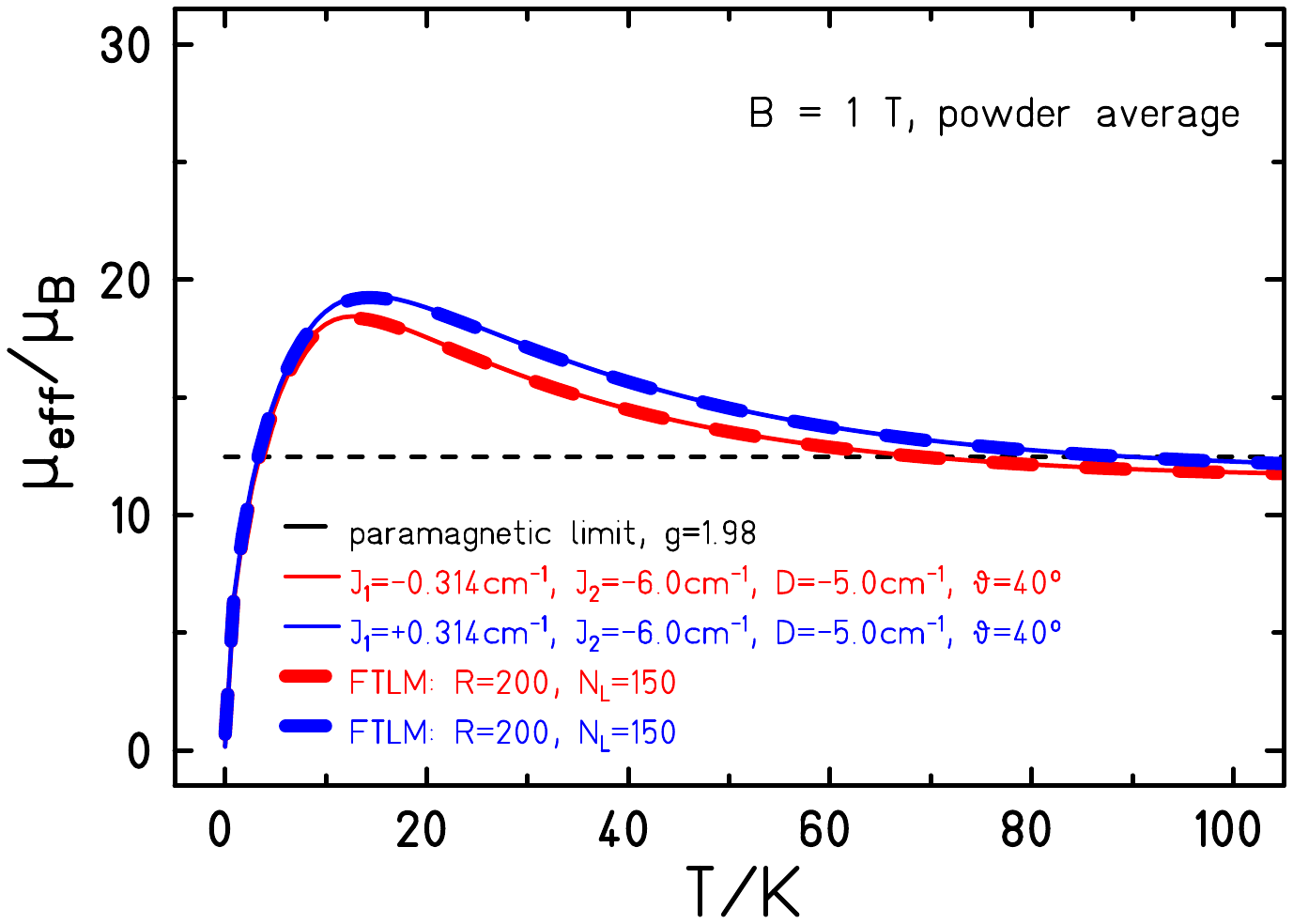}
\caption{Effective magnetic moment of model system \textbf{M2}:
  moment along $z$-direction (top), powder averaged
  moment (bottom). The solid curves show the result of full matrix
  diagonalization, the dashed ones the result of FTLM.} 
\label{tlmm-f-4}
\end{figure}

Model system \textbf{M2} is inspired by
Mn$_6^{\text{III}}$Cr$^{\text{III}}$  molecules \cite{HHK:IC12},
where 6 spins $s=2$ are arranged in two  
equilateral triangles with fictitious nearest neighbor exchange
interaction $J_1=\pm 0.314$~cm$^{-1}$ and a seventh central ion
with $s=3/2$ connected with $J_2=-6.0$~cm$^{-1}$ to all other
spins. The single-ion anisotropy tensors with $D_i=-5.0$~cm$^{-1}$,
$E_i=0$ for the Mn spins are the same as in \textbf{M1},
compare \figref{tlmm-f-1}; for chromium we choose $D_7=E_7=0$. 
The size of the Hilbert space is
62.500. The exact evaluation of a powder average with 25
orientations for just one field value needs about a week,
thereby gobbling up 46~GB of RAM. The corresponding FTLM
simulations need less than one hour on a simple notebook. Figure
\xref{tlmm-f-4} displays again the effective magnetic moment
along $z$-direction (top) and as a powder average (bottom). The
solid curves show the result of full matrix diagonalization, the
dashed ones the result of FTLM. Again, the accuracy is
astonishing. 

This success is very encouraging for two reasons: With a
numerically exact diagonalization it is virtually 
impossible to perform large parameter searches for molecules as
big as \textbf{M2}, while using FTLM it becomes feasible. In
addition, one can now numerically investigate much
larger anisotropic spin systems both as function of temperature
and field with high accuracy. This will be demonstrated in the
following.

Several of the most interesting molecular magnets, i.e. magnetic
molecules possessing a magnetic hysteresis, are of bigger
size. The famous Mn$_{12}$-acetate molecules contain 12
manganese ions of two valencies (8 Mn$^{3+}$ ions with $s=2$ and
4 Mn$^{4+}$ ions with $s=3/2$)
\cite{Lis:ACB80,STS:JACS93,SGC:Nat93,TLB:Nature96,LTB:JAP97,ThB:JLTP98,CGJ:PRL00}. 
The resulting dimension of the Hilbert space assumes exactly
100,000,000. It was so far impossible to treat such a molecule
on the basis of a full spin Hamiltonian including single-ion
anisotropy. First attempts have been made using Lanczos
procedures for low-lying states as well as high-temperature
series expansions 
\cite{RJS:PRB02,CSO:PRB04}. 
Other spin systems of similar size are the mixed valent
Mn$_{12}$ ring of Christou \cite{STS:JACS93} as well as the
mono-valent Mn$_{12}^{\text{III}}$ ring of Brechin
\cite{SFR:CAEJ14}.

Since it is our aim to demonstrate that the extension of FTLM
towards anisotropic spin systems works we consider a fictitious  
Mn$_{12}^{\text{III}}$ ring with uniaxial anisotropy. 
This has the advantage that at least for magnetic fields along
the anisotropy axis we can compare FTLM codes employing
$\op{S}^z$ symmetry with the new method. The Hamiltonian thus
can be written as 
\begin{eqnarray}
\label{E-4-1}
\op{H}
&=&
-
2\;J\;
\sum_{i}\;
\op{\vec{s}}_i \cdot \op{\vec{s}}_{i+1}
+
D\;
\sum_{i}\;
\left(\op{\vec{s}}_i^z\right)^2
\\
&&+
g\, \mu_B\, \vec{B}\cdot\op{\vec{S}}
\nonumber
\ .
\end{eqnarray}
We investigated magnetic observables for several orientations of
the external magnetic field for the following parameters of the
spin system: $J=\pm 3.0$~cm$^{-1}$, $D=-1.8$~cm$^{-1}$, and
$g=1.98$. The dimension of the Hilbert space is 244,140,625.

\begin{figure}[ht!]
\centering
\includegraphics*[clip,width=70mm]{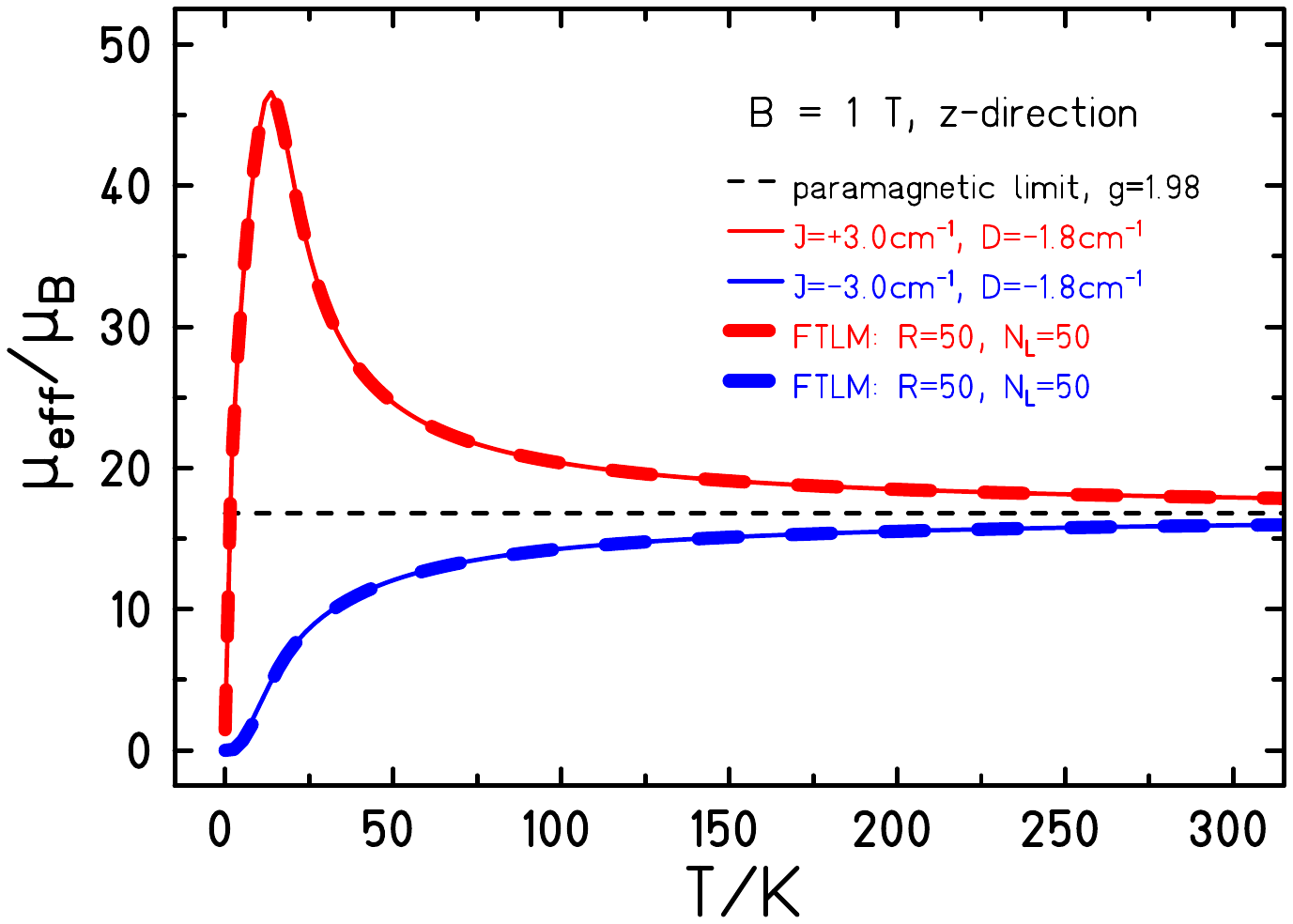}
\caption{Effective magnetic moment of a fictitious
   Mn$_{12}^{\text{III}}$: The solid
   curves are produced with a FTLM code employing spin
   rotational symmetry about the $z$-axis ($R=20, N_L=120$),
   whereas the broken curves are calculated with the
   method proposed in this article.}   
\label{tlmm-f-5}
\end{figure}

The case where the applied field points along the
uniaxial anisotropies can be treated with a FTLM code where the
Hilbert space is decomposed into subspaces ${\mathcal H}(M)$,
compare \cite{ScW:EPJB10,ScH:EPJB13}. The resulting effective magnetic
moment is depicted
by solid curves in \figref{tlmm-f-5}. The effective moment for a 
ferromagnetic coupling shows the typical maximum, whereas for
the antiferromagnetic case, the effective moment simply rises
with increasing temperature. The dashed curves present the
results of our proposed FTLM. The agreement is again very good. 

\begin{figure}[ht!]
\centering
\includegraphics*[clip,width=70mm]{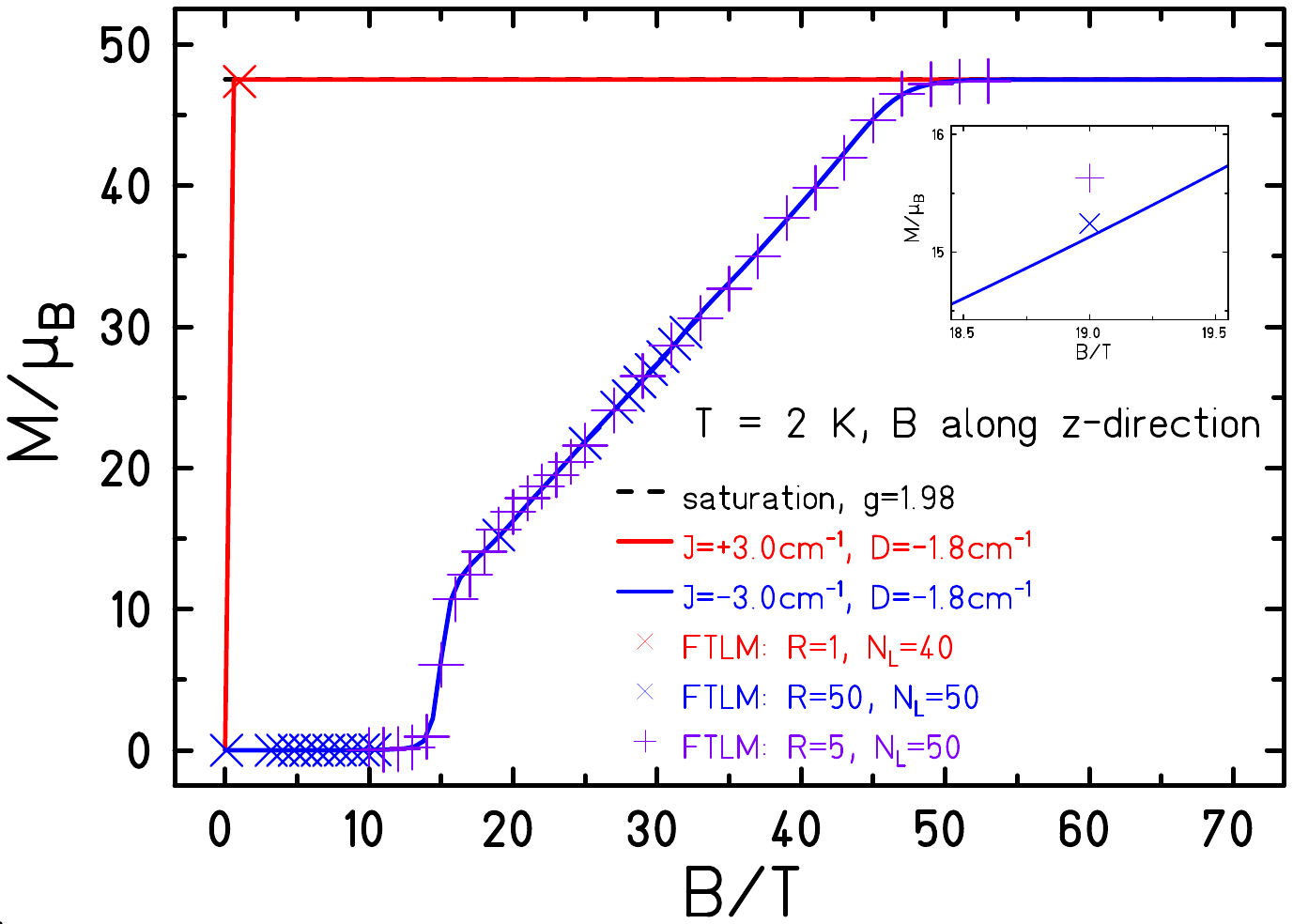}

\includegraphics*[clip,width=70mm]{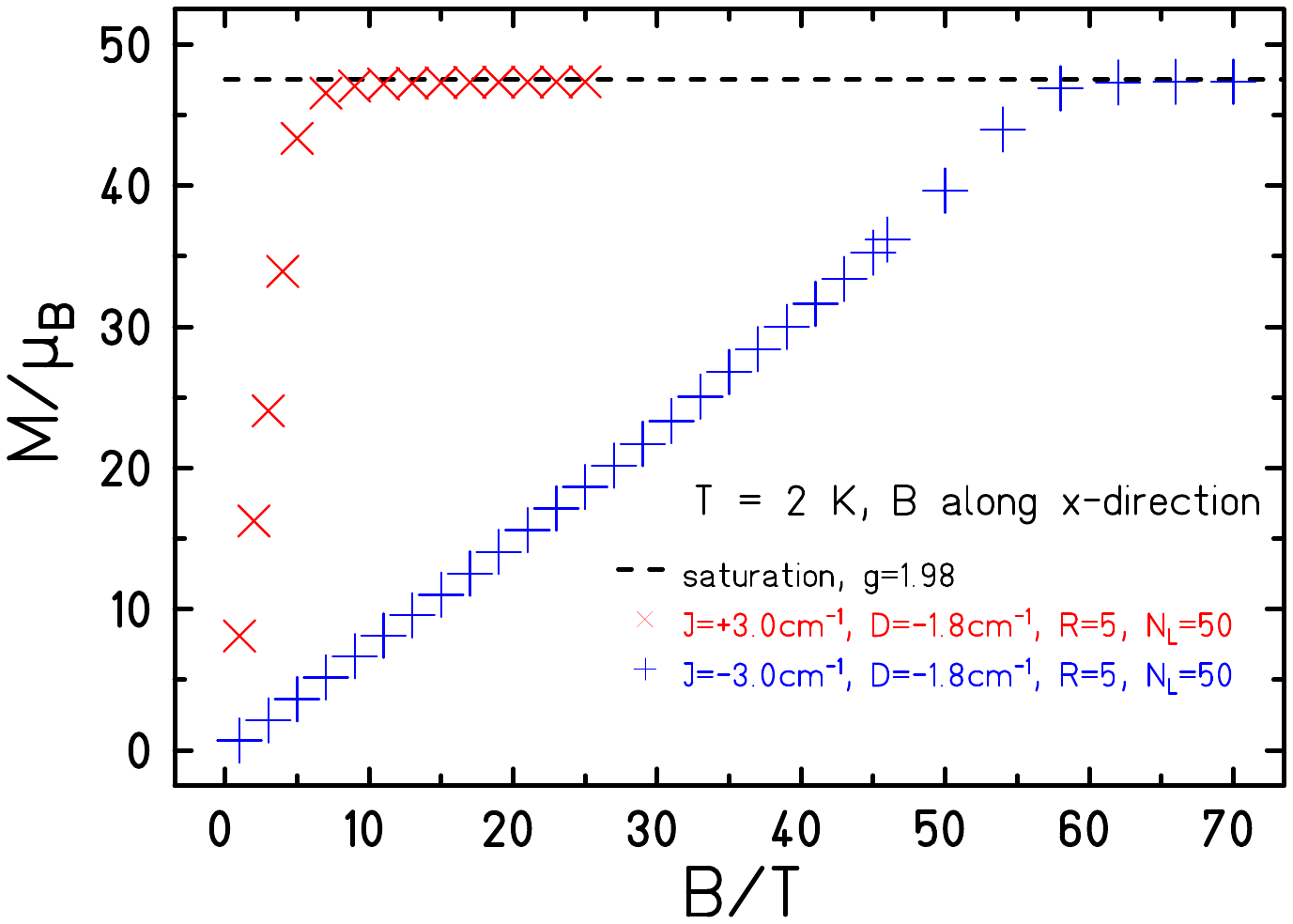}
\caption{Magnetization of a fictitious Mn$_{12}^{\text{III}}$:
   $B$ along $z$-direction (top), $B$ along $x$-direction
   (bottom). The solid curves are produced with a FTLM code 
   employing spin    rotational symmetry about the $z$-axis
   ($R=100, N_L=120$). The symbols depict the magnetization
   calculated with the method proposed in this article for
   various numbers of random vectors. The inset shows for a
   single data point with larger deviation how the numerical
   result improves when using 50 instead of 5 random vectors.}
\label{tlmm-f-6}
\end{figure}

The magnetization is shown in \figref{tlmm-f-6} for $B$ pointing
along $z$- and $x$-direction. In the ferromagnetic case the
magnetization for a field along $z$-direction closely follows the
Brillouin function of a total spin $S=24$, and thus
immidiately jumps to saturation. In the antiferromagnetic case
the staircase bahavior of a pure Heisenberg ring is smeared out
due to anisotropy (and temperature). Both functions are rather
well reproduced by the proposed FTLM. Nevertheless, now the
evaluations need some time: a single data point with $R=5$ and
$N_L=50$ needs about eight hours on 128 cores of our local SMP
machine. Therefore, we evaluated only one data point for the
ferromagnetic case, but several for the antiferromagnetic case,
compare \figref{tlmm-f-6} (top). For the antiferromagnetic case
we again investigated the influence of the number of random
vectors $R$. We find that for $T=2$~K, which is of the order of
the parameters of the Hamiltonian, five random vectors (and
their symmetry related counterparts) are sufficient for most of
the curve. Only around $B=20$~T, where we observe a small
deviation, we find that an increased number of random vectors
($R=50$) is necessary to yield a good approximation, compare
inset of \figref{tlmm-f-6} (top). As expected for a
Monte-Carlo-type procedure the 
deviations are about $1/\sqrt{10}$ times smaller for 10 times
more random vectors. 

In the case where the external field 
is applied perpendicular to the easy axes, \figref{tlmm-f-6}
(bottom), the magnetization rises more slowly both in the
ferromagnetic as well as in the antiferromagnetic
case. Observable such as this one cannot be evaluated with any other
method (with the same accuracy).

\section{Summary and Outlook}
\label{sec-5}

After countless efforts to develop numerical
strategies that rest on  the
symmetries of a quantum spin problem
\cite{DGP:IC93,BCC:IC99,Wal:PRB00,BOS:TMP06,ScS:PRB09,ScS:IRPC10},
nowadays approximate methods such as Quantum Monte Carlo
\cite{SaK:PRB91,San:PRB99,EnL:PRB06}, Density 
Matrix Renormalization Group Methods
\cite{Whi:PRB93,Sch:RMP05} and in particular Krylov
space based methods such as FTLM produce approximate results of
unprecedented accuracy. The latter is in particlar encouraging since
Lanczos methods are very easy to program whereas irreducible
representations of SU(2) combined with point groups have
been mastered by only a rather small group of experts.
In addition they don't suffer from restrictions such as the
negative sign problem that Quantum Monte Carlo faces for
frustrated spin systems.

We thus hope that we could convince the reader that the
Finite-Temperature Lanczos Method is capable of evaluating the
thermal properties of large quantum spin systems even if they 
lack the $\op{S}^z$ symmetry.

\section*{Acknowledgment}

This work was supported by the German Science Foundation (DFG
SCHN 615/15-1).
Computing time at the Leibniz
Computing Center in Garching is also gratefully
acknowledged. 


\end{document}